# An Agent-Based Discrete Event Simulation of Teleoperated Driving in Freight Transport Operations: The Fleet Sizing Problem

Bahman Madadi[a]

Ali Nadi[a*]

Gonçalo Homem de Almeida Correia[a]

Thierry Verduijn [b]

Lóránt Tavasszy[a]

[a] Delft University of Technology, Delft, The Netherlands

[b] Rotterdam University of Applied Sciences, The Netherlands

* Corresponding author: a.nadinajafabadi@tudelft.nl

2023

## ABSTRACT

Teleoperated or remote-controlled driving complements automated driving and acts as transitional technology towards full automation. An economic advantage of teleoperated driving in logistics operations lies in managing fleets with fewer teleoperators compared to vehicles with in-vehicle drivers. This alleviates growing truck driver shortage problems in the logistics industry and save costs. However, a trade-off exists between the teleoperator-to-vehicle ratio and the service level of teleoperation. This study designs a simulation framework to explore this trade-off generating multiple performance indicators as proxies for teleoperation service level. By applying the framework, we identify factors influencing the trade-off and optimal teleoperator-to-vehicle ratios under different scenarios. Our case study on road freight tours in The Netherlands reveals that for any operational settings, a teleoperation-to-vehicle ratio below one can manage all freight truck tours without delay, while one represents the current situation. The minimum teleoperator-to-vehicle ratio for zero-delay operations is never above 0.6, implying a minimum of 40% teleoperation labor cost saving. For operations where a small delay is allowed, teleoperator-to-vehicle ratios as low as 0.4 are shown to be feasible, which indicates potential savings of up to 60%. This confirms great promise for a positive business case for the teleoperated driving as a service.

# 1   Introduction

Connected and automated driving is expected to revolutionize transportation and logistics by providing benefits such as safety, traffic efficiency, comfort, and reducing emissions as well as enabling novel concepts such as robotaxis, car-sharing, and truck platooning [1]. Recent developments in vehicle and communication technologies have enabled connected and automated driving in certain controlled environments (e.g., driving on motorways under normal weather conditions). However, some challenges for enabling connected and automated driving in all driving domains and under all conditions remain unresolved.

According to [2], there are six levels of vehicle automation. Driving automation systems at level-0, level-1, and level-2 provide the driver with longitudinal and lateral support (i.e., emerging braking, adaptive cruise control, and lane keeping). Such technologies are available on some vehicles currently sold on the market and are rapidly becoming more commonplace. They can be classified as "hand- and/or feet off driving". At level-3, automated driving systems monitor the environment and execute driving tasks on certain operating design domains (e.g., driving in motorways), allowing the drivers to avert their attention from driving tasks while being ready to take back control in case of a failure in the automated driving system. This level is also referred to as "eyes-off driving". Level-4 automated driving systems, also referred to as "mind-off driving", are expected to handle the fail-safe situation autonomously; however, within a limited operating design domain. Therefore, level-3 and level-4 automated driving systems can only be activated on specific road segments and specific conditions. Finally, level 5 refers to fully autonomous vehicles with unlimited operating design domains. This last level of automation signals a major evolution in the prospect of mobility, but it is not expected soon [3].

The analyses of automated driving system disengagements occurring during the automated vehicle tests in the United States indicate that the existing vehicles are not capable of performing all dynamic driving tasks reliably and flawlessly in all conditions, particularly in complex urban environments [4], [5], [6], [7]. Some studies have proposed solutions to remedy this issue by adjusting the infrastructure and whitelisting to utilize automated driving on selected roads [8], [9], [10] or via dedicated lanes for automated vehicles [11], [12]. However, these solutions can be costly.

Teleoperated driving (TOD) or remote-controlled driving could be complementary to automated driving (e.g., a teleoperator taking control in particularly complex driving situations) as well as a transition technology to fully automated driving [13], [14]. Modern teleoperation has been in use since the 1940's and has been applied in various fields, such as space exploration, military operations, mining, surgery, and port operations [15], [16].

In general, teleoperation (TO) refers to a system where a human being controls a robot from a distance. Any teleoperated system is defined by three main elements: the robot, the operator interface, and the communications link [17]. The robot, which is the vehicle in the case of TOD, integrates mechanical and electronic components. Its design varies over operating environments and application domains. In TOD, the vehicle is equipped with cameras and sensors (e.g., radar and lidar) to monitor its environment, and possibly onboard processors and software to analyze and perceive the environment. The operator interface generally consists of displays to show the information gathered by the robot's sensors and input devices for the operator to enter commands and execute control over the robot. The communication link provides the means for two-way communication to allow the flow of information between the robot and the operator (i.e., information gathered by the robot's sensors and the command entries by the operator). For TOD, general definitions, as well as TOD system design and architecture, are provided by [14], [15], [16], [17], [18]. Moreover, several simulation tools for teleoperated driving have been proposed, and tested [19], [20].

Teleoperated driving is expected to have major implications for logistics and fleet operations as well. It is suggested in [21] that teleoperated taxi fleets could revolutionize urban mobility by offering a cost-effective and safe door-to-door transportation service. The authors use an empirical evaluation to conclude that the implementation of the service can reduce the number of drivers by up to 27%. The operational performance of fleets of teleoperated vehicles is explored by [22]. The authors assumed that a team of teleoperators would be responsible for monitoring a large fleet of automated vehicles and would take control of the vehicle upon request by the vehicles' automated driving system. Such concepts are relevant when the teleoperated vehicles are level-4 automated vehicles.

Hjelt in [23] studied the total cost of ownership of autonomously operated buses at autonomy level 4 and 5 supported by remote operators. Teleoperation could also tackle the critical first and last mile in passenger car and truck platooning, significantly enhancing the chances of bringing this to reality [13], [24], which can reduce logistics or fleet operations costs and environmental impacts.

The adoption of teleoperated driving could also help in tackling growing operator shortages in the logistics industry. When it comes to truck drivers, demand in the Netherlands and Belgium is growing steadily while the supply is lagging due to uncompetitive labor conditions, long working hours, and prolonged periods away from home. This has led to persistent shortages of truck drivers [25], [26], [27]. Teleoperation has the potential to solve these problems by converting truck drivers into teleoperators, thereby eliminating the need for difficult working conditions, and working away from home.

The main economic benefit of teleoperation for logistics operators is expected to be lower labor requirements, especially where the labor shortage is an economic barrier. Because with teleoperation, logistic service providers can keep a fleet of vehicles operational with a lower number of teleoperators compared to the number of vehicles. However, the viability of teleoperation in the freight transport context depends on the teleoperator-to-vehicle (TO/V) ratios.

Therefore, the main aim of this study is to explore the factors influencing the teleoperation labor cost as well as the trade-offs between the teleoperation service level and the fleet size. Using a case study of road freight tours in The Netherlands, we attempt to determine the optimal teleoperator-to-vehicle ratios in different logistics operations given a certain level of service and to identify factors that have an impact on the optimal fleet size and consequently the teleoperation labor cost under different deployment scenarios. Since teleoperation is not yet sufficiently operational to be tested on the field, we develop a simulation framework to examine this concept under four scenarios.

The remaining of this manuscript is organized as follows. Section 2 includes a description of our methodology and data collection. Section 3 describes the experimental setup and the case studies. Section 4 presents the results of the case studies as well as the analysis of the results. Finally, the last section contains the concluding remarks.

## 2 Methodology

The main aim of this study is to define the optimal teleoperator-to-vehicle ratio for a given teleoperation service level in logistics operations. Therefore, we design a simulation tool that replicates the teleoperator allocation and queuing processes of a teleoperation center providing teleoperated driving as a service to logistics service providers. The teleoperated driving simulation requires road freight tours as input to represent the teleoperation demand. Therefore, a simulation framework that replicates individual firms' logistics decision-making taking place at the level of shipments is required. To accurately estimate the teleoperation demand, we resort to MASS-GT, a multi-agent simulation system for goods transport that simulates the logistics decision-making behind freight transportation demand [28]. It also simulates a large variety of decision-makers and choices at the level of individual firms taking into account agent-specific costs and constraints. The MASS-GT simulation model has been used widely in many real-world applications [29], [30], [31], [32]. Our methodology integrates the teleoperation simulation tool developed in this study with MASS-GT to generate accurate estimations of road freight tours, use the generated tours as demand for teleoperation within the teleoperation simulation, and provide indicators for system performance under various scenarios.

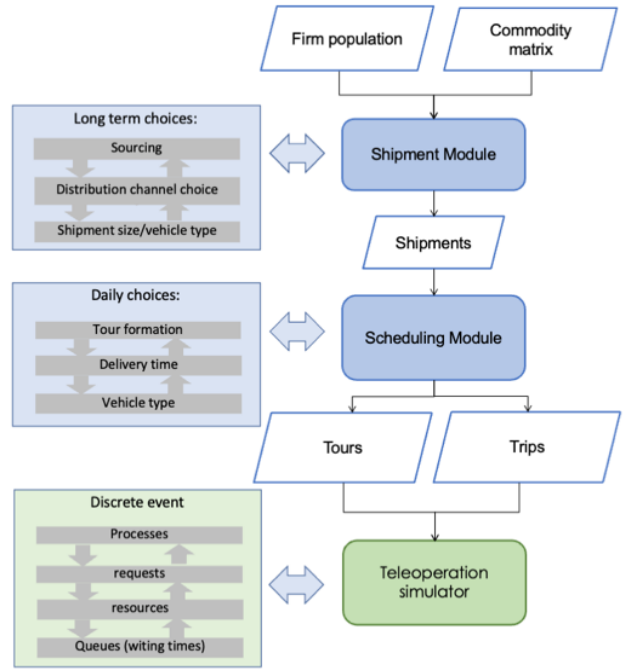

*Figure 1 Extended MASS-GT with Teleoperation simulation module.*

Our methodology contributes to the development of MASS-GT by adding another building block to it that enables it to evaluate futuristic and innovative solutions such as teleoperated driving, yet when the tour data is available, the teleoperation simulation module can be used as a stand-alone simulation application as well. Fig 1 shows the building blocks of the simulation and how the proposed teleoperation simulation is integrated with it. The heart of MASS-GT encompasses two levels of logistic decision-making: long-term tactical choices, which are simulated in the shipment module, and short-term tactical choices, which are simulated in the scheduling module. These two fundamental modules simulate freight transport demand at the shipment level. The output of this simulation is a set of scheduled tours that transport simulated shipments from their origins to their destinations.

As shown in Fig 1, the teleoperation simulator module requires trips and tours data, which are generated within the freight demand simulator by running the shipment module and the scheduling module, respectively.

## 2.1 Shipment module

The shipment module synthesizes a set of shipments that are transported in the study area. To create this set of shipments, an event-based simulation is used for the four logistic processes: 1) producer selection; 2) distribution channel choices; 3) shipment size & vehicle type choice; and 4) desired delivery time [31], [33].

## 2.2 Scheduling module

The second is the scheduling module that simulates the formation of tours, chooses the time for each tour based on the desired delivery times, and optimizes the vehicle type choices. Time-of-day decisions are simulated both in the Shipment module, and the Scheduling module. In the shipment module, first, a choice for the desired delivery time for each shipment is determined. In the scheduling module, the desired delivery time is taken into account in the selection of shipments that are considered for consolidation, the delivery order, and the tour departure time. For further information about this procedure, we refer the readers to [33], [34].

The end product after running the shipment and scheduling module is a set of trips shaped in the form of tours for each firm. These tour activities are inputs for the teleoperation simulator.

### 2.3 Teleoperation simulator

The teleoperation simulator begins by processing tour activity sequences that have been simulated by MASS-GT in the study area. To determine a representative sample of tours, we make an assumption about the market penetration of the teleoperation system and then select a set of tours at random. Since the main purpose of the simulation study is to explore the optimal teleoperator-to-vehicle ratios given a certain level of service for teleoperation, the following assumptions are made.

- The simulation focuses on the driving tasks performed by teleoperators.
- Driver responsibilities other than driving are not considered in the simulation.
- Company operations and activities will stay the same with teleoperation.
- The market penetration rate of teleoperation is 1% of all freight transport activities simulated for a day.

Given these assumptions, we propose a simulation procedure of teleoperation which is explained in the next subsection.

### 2.4 Simulation procedure

Our simulation is built upon Discrete Event Simulation (DES) techniques. The main components required for the discrete event simulation are the system, the entities (trucks), and the resources or servers (teleoperators). The system is the process or phenomenon being modeled by the simulation. In our case, the system is teleoperation of transporting goods from their origins to their destinations. In this system, we can identify two types of agents, namely, trucks and teleoperators.

Each truck has a sequence of activities that defines its state. These activities have certain characteristics such as start, duration, and end, which are derived from MASS-GT tour data. During the simulation, the state of each truck can be updated as:

- Idle (buffer): when a truck is loading or unloading goods.
- In Queue: when there is no teleoperator available and the vehicle should wait in a queue.
- Takeover: when a teleoperator accepted the truck request and is taking over the control.
- Teleoperated: when the trip starts and the teleoperator is operating the vehicle.
- Signed off: when the vehicle has finished its tour.

Similarly, each teleoperator has a certain state variable which can be updated as follows:

- Idle: When a teleoperator is free and waiting for a request.
- Busy: when the teleoperator is operating a truck.
- Resting: when the teleoperator has reached its maximum allowed hours of teleoperation.
- Takeover: when a teleoperator is taking over the control of a truck.

The DES is also constituted by Events which are occurrences that trigger changes in the state of the entities and resources; the Time/clock that represents the passage of time through a series of discrete steps; and the Queues, which are places where entities can wait to be served by teleoperators. Since a limited number of teleoperators (i.e., resources) are available in each scenario, a queue might grow for teleoperators. This queue is modeled as a stochastic process based on the first in first out (FIFO) rule. Although the arrival process of this queueing model is determined based on the departure time of the trips (trucks request teleoperators at their departure times), this process is still stochastic. This is because the departure time of the tours in MASS-GT is based on the Monte-Carlo simulation of a discrete choice model that is calibrated based on real-world data of tour schedules [31], [34]. This queueing model is also a multi-server process as we can have m number of teleoperators in the system (see Fig 2). The service rate and process time of each teleoperator depend on the duration of the trips that they are teleoperating and takeover times.

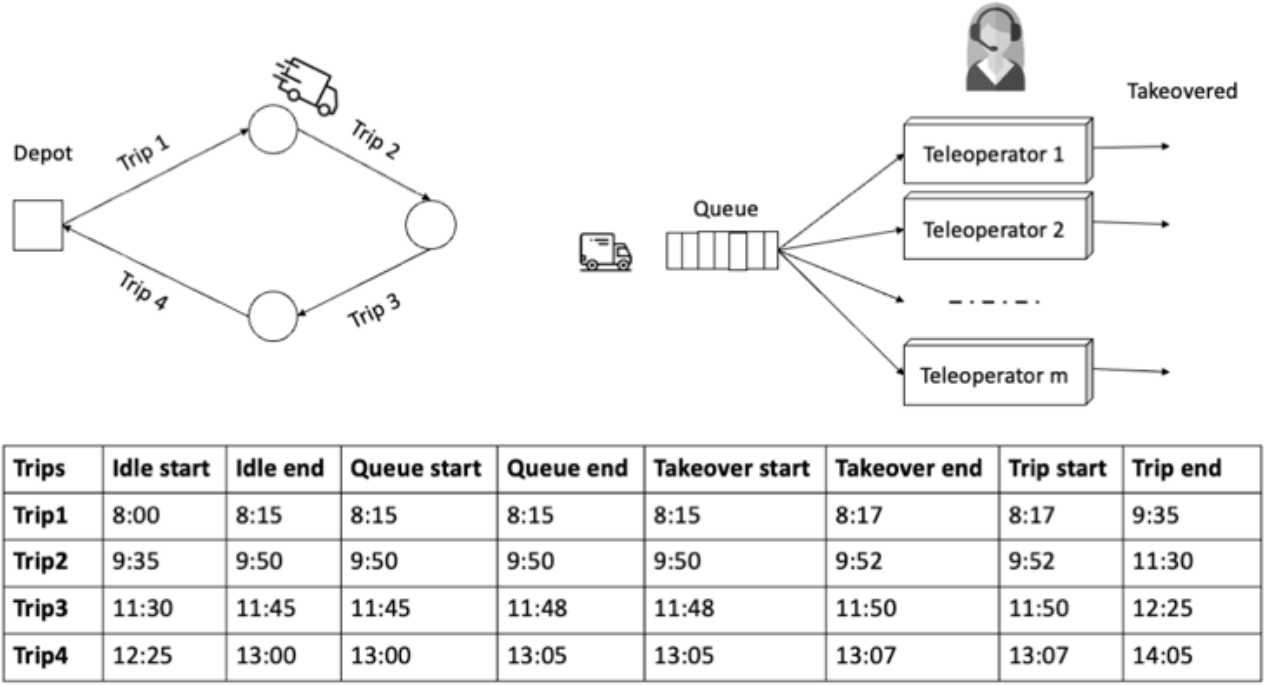

*Figure 2  Multi-server queue procedure and update of vehicle state characteristics.*

Finally, the last component of the DES is the Statistics which are used to record information about the system during and after each run. Examples of statistics include the number of entities in the system, the average waiting time for an entity, and the utilization of resources. These components work together (as shown in Fig 3) to create a model of the system being simulated, which allows us to evaluate different scenarios and identify potential improvements.

After the initialization of the simulation parameters, the clock starts. For all trucks in the pool of vehicles, we select the first event in the activity list of the vehicle. If the vehicle state is Idle and the clock of the simulation is equal to the end of the Idle time, then the vehicle sends a takeover request to the teleoperator center. If all teleoperators are busy, then the truck should wait in the queue and the queue processor adds to the length of the queue. Otherwise, a teleoperator will take over the vehicle and the status of the vehicle and teleoperator will be updated. Then, the truck's trip is simulated from the origin to the destination. When the trip ends, the teleoperator status will be updated to Resting if the teleoperator has been busy for a certain number of hours. If the tour has ended, then the vehicle status will be updated to signed-off, otherwise, the next event in the vehicle activity list is triggered. Fig. 3 illustrates this simulation procedure in more detail.

### 2.5    Key performance indicators

In order to evaluate teleoperation scenarios, one needs to define key performance indicators (KPI).

*Waiting time per vehicle*

The first KPI we define is the average waiting time ($wt$) per vehicle where k represents each truck ($K$ being the number of trucks), $T^k$ represents the total number of trips in a tour that vehicle k must be driven. $w_t^k$ is the time that truck $k$ must wait in a queue before starting trip $t \in T^k$.

$$wt_k = \frac{1}{K} \sum_{k=1}^{K} \sum_{t=1}^{T} w_t^k \qquad (1)$$

*Waiting time per vehicle in the queue*

This KPI indicates the duration of waiting each time a vehicle enters a queue. Therefore, it equals cumulative waiting duration for all vehicles divided by the total number of times that vehicles enter the queue ($NQ$).

$$wt_Q = \frac{1}{NQ} \sum_{k=1}^{K} \sum_{t=1}^{T} w_t^k \qquad (2)$$

*Vehicle utilization*

Another KPI used in this simulation is vehicle utilization as defined below.

$$Util_k = \frac{\sum_{t=1}^{T} TT_t^k}{ST} \quad \forall \ k \in K \qquad (3)$$

where $Util_k$, utilization of vehicle $k$, is the amount of time each vehicle is moving ($TT_t^k$ is the travel time of trip $t$ in the tour of vehicle $k$) divided by the simulation time $ST$.

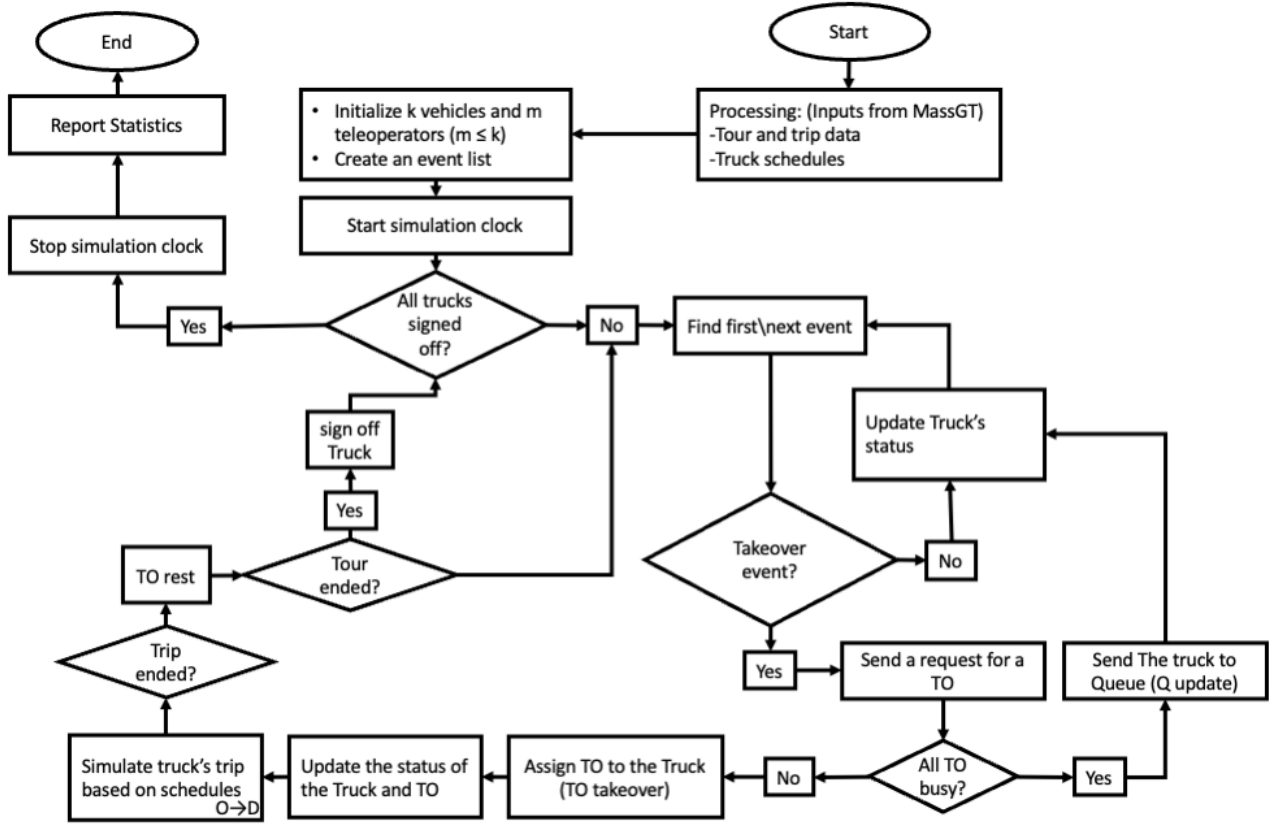

*Figure 3  Teleoperation simulation procedure*

*Teleoperator utilization*

Likewise, the utilization of teleoperators is defined as follows.

$$Util_{to} = \frac{\sum_{t=1}^{T} TT_t^{to}}{ST} \qquad \forall \ to \in TO \qquad (4)$$

This is the amount of time each teleoperator $to$ in the set of teleoperators ($TO$) is busy driving a vehicle in trip $t$, including rest and takeover times, divided by the total simulation time.

*Makespan*

Another important KPI that allows us to compare scenarios is the makespan which is the total time required in the simulation to complete all tours.

$$Ms = \sum_{k \in K} \sum_{t \in T} (TT_t^k + su_t^k + w_t^k) \tag{5}$$

In Equation 5, $su_t^k$ is the setup time (takeover time) of vehicle $k$ in trip $t$.

*Tour completion rate*

The makespan indicator helps us to define two important indicators that allow comparing the different settings of the teleoperation system with the baseline scenario. The first indicator is Tour Completion Rate (TCR) which is the number of completed tours at the baseline makespan $CT_{Ms_b}$ divided by the total complete tours at the scenario makespan $CT_{Ms}$.

$$TCR = \frac{CT_{Ms_b}}{CT_{Ms}} \tag{6}$$

*Distance completion rate*

There are some cases where tours are just remarkably close to being completed at the baseline makespan. Then the TCR cannot consider them as completed tours and hence shows a lower level of service for the system. Therefore, we defined the Distance Completion Rate (DCR) which is the total completed tour distance $CTD_{Ms_b}$ at the baseline makespan divided by the total planned tour distance $CTD_{Ms}$. This measure provides a more accurate picture of the tour progress at baseline makespan.

$$DCR = \frac{CTD_{Ms_b}}{CTD_{Ms}} \tag{7}$$

*Delay*

Finally, we define delay rate as the normalized difference between the total simulation duration of the scenario and the total duration of the baseline.

$$D = \frac{CT_{Ms} - CT_{Ms_b}}{CT_{Ms_b}} \tag{8}$$

To account for the uncertainty of the simulation results as well as the stochastic variations within runs, we report the average, standard deviation, minimum, median, and maximum of the KPIs derived from 5 replications of the simulation for each scenario.

## 3 Numerical experiments in the South Holland case study

To conduct a simulation experiment with a teleoperation simulator, the MassGT needs to be executed first, as shown in Figure 1. The shipment module relies on both aggregated and disaggregated data from various sources. The Netherlands Statistics Bureau provides the primary source of data, using an XML interface to automatically extract microdata from the Transport Management Systems (TMS) of transport companies.

The data includes details on the vehicle, route, commodity type, weight, and loading and unloading locations, but not on the shippers and receivers of goods. To determine these locations, aggregate statistics from the Netherlands firm registration data (ABR) are used. Additionally, data on distribution centers (DC) and transshipment terminals (TT) are obtained from Rijkswaterstaat and contain information on their addresses, sizes, and sectors. The CBS trip diaries are enriched with this additional location information [35]. These data provide MASS-GT with a firm population. Finally, the regional commodity flow data are derived from the Dutch strategic freight model "BasGoed" [36]. In the next section, descriptive statistics of the input and output of the MASS-GT simulator are provided in the study area. These data provide MASS-GT with inputs as well as ground truth for the calibration of the choice models. We refer readers to [28] and [33] for further information about MASS-GT choice models and their calibration.

### 3.1 Descriptive statistics of the data

The study area of the current research is the province of South Holland and all freight transportation that takes place within, to, and from this study area. The simulation and its underlying behavioral models are empirically calibrated and validated based on large observed real-world logistics data. This data is available from an automated data collection of truck trip diaries among a sample of Dutch truck owners that is being collected by the Central Bureau of Statistics Netherlands (CBS) [28], [30], [33], [35]. The data includes information on the vehicles' characteristics, e.g., vehicle types and capacity, tours (loading and unloading locations and time), and shipment information within these tours, e.g., commodity type and size.

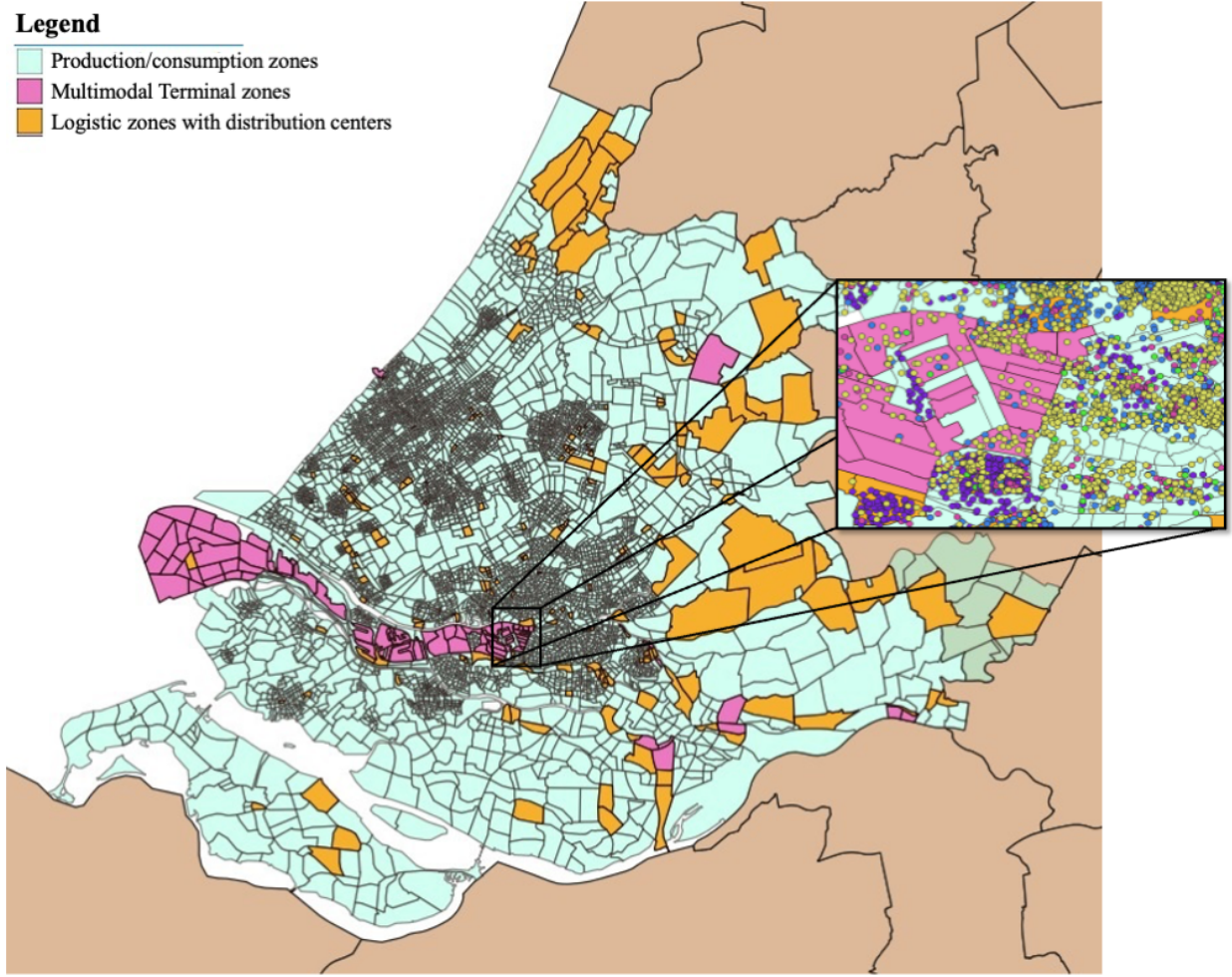

*Figure 4 Study area with the population density of logistic activities*

The firm population is generated and scattered across the Netherlands. Based on the activity type of firms and their density, the zones can be classified into Production or Consumption zones, Multimodal terminal zones, and zones with logistic activities like Distribution centers (see Fig. 4).

After running MASS-GT, the total number of tours simulated for an average day is just above 124,000 tours with more than 450,000 trips. For each simulation scenario, we filter all tours to select the ones that fall within the simulation time window according to the corresponding scenario (simulation start time and duration are scenario parameters). For each simulation replica within each scenario, we select 1% of the tours to be controlled by teleoperation (see Fig. 5).

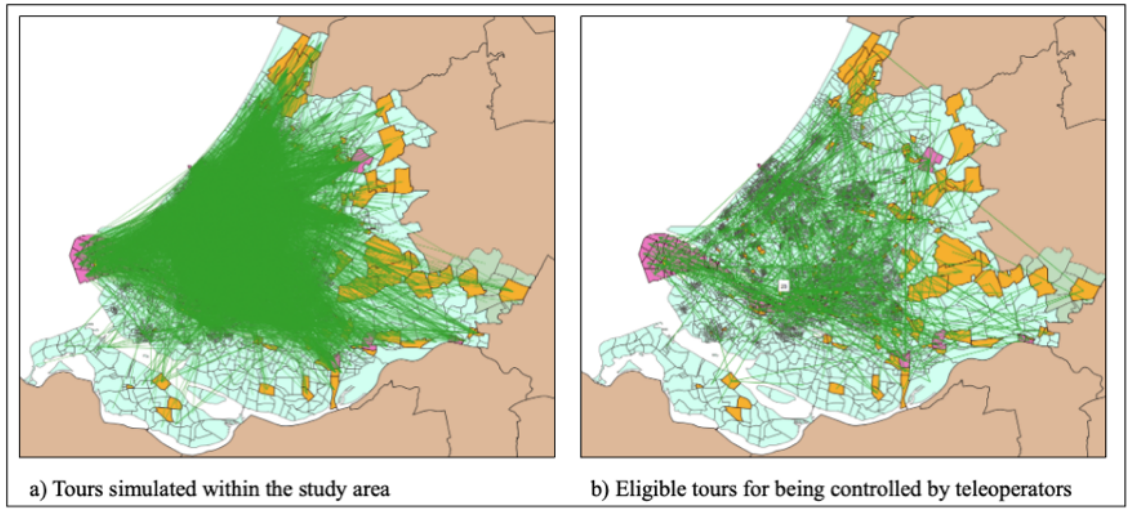

*Figure 5  Simulated tours in the study area and the eligible tours for teleoperation*

Table 1 shows a descriptive summary of a sample of eligible tours for the base case scenario. These simulated tours are validated using the CBS XML data. We refer readers to [34] for empirical evidence on the validity of the MASS-GT freight tour simulation.

*Table 1  General descriptive statistics*

| Indicators | Values |
| --- | --- |
| Number of tours | 454,170 |
| The average number of trips per tour | 4 |
| Total number of tours | 124,436 |
| The average tour duration (hours) | 7.37 |
| The average trip duration per tour (hours) | 3.71 |
| Number of teleoperated tours | 1244 |
| Number of teleoperated trips | 5326 |

### 3.2  Experimental setup

To evaluate the efficiency of teleoperation and potentially reduce the number of required human resources, we have designed various scenarios for this study. The teleoperator-to-vehicle ratio is the primary parameter in defining these scenarios. Although lower ratios are more cost-effective, they may result in longer waiting times for vehicles and lower levels of service due to queuing for teleoperators. To strike a balance between teleoperator costs and service level, we used a grid of teleoperator-to-vehicle ratios ranging from 0.3 to 1. In our initial interviews with experts, we found that a short setup (takeover) time for teleoperators was necessary for safety reasons and to allow for situational awareness before driving. Therefore, we studied scenarios with one-, two- and three-minute setup times and included scenarios without this setup time for comparison purposes. One of the significant factors in the economic feasibility of teleoperated driving is the number of teleoperators required for a fleet of vehicles. Although reducing the number of teleoperators can decrease labor costs and improve utilization, it can also create inefficiencies in logistics facilities due to longer waiting times for trucks and occupying docks unnecessarily. The level of service of teleoperated fleet operations can be determined by the percentage of trips in which the waiting time of teleoperated trucks does not exceed the agreed maximum waiting time. However, this maximum threshold depends on the policies and level of resiliency of the freight transport system and needs further studies. We, therefore, use the waiting time itself as the proxy for the level of service of the teleoperation. Table 2 summarizes the parameters and variables in the simulation scenarios.

In this futuristic experiment, we assume that the market penetration of teleoperation will not exceed 1%. This is due to the requirement of advanced communication technologies that must be installed on the vehicles, which may not yet be affordable in the market, especially for small sectors. The willingness of the transportation industries to enhance their fleets with such technologies is still a topic for research and is beyond the scope of this study.

*Table 2  Scenario variables and parameters*

| Scenario variable | Values | Description |
| --- | --- | --- |
| Simulation start time | 0:00 – 05:00 – 8:00 | This defines the different settings for the start time of teleoperation within a day. |
| Simulation duration | 9 hours – 24 hours | This defines the working shift of teleoperation within a day. |
| Ratio of teleoperators/Vehicle | 0.3-1 | This allows us to investigate the level of service for different operator-to-vehicle ratio |
| Take over time | 0, 1,2, 3 (minutes) | The time each teleoperator require to take over a vehicle |
| **Parameter** | **Value** | **Description** |
| Number of replications | 5 | The number of simulations run for each scenario |
| Market penetration | 0.01 (1%) | The percentage of transport sectors that are eligible for teleoperation |
| Maximum allowable teleoperation | 4.5 hours | For safety and health-care reasons, each teleoperator is only allowed to control a vehicle for a maximum amount of 4.5 hours. |
| Rest time | Short: 10 minutes<br>Long: 45 minutes | Rest time for teleoperator after a teleoperation task. |

We have defined three different simulation start times which can be practically interpreted as the time that the teleoperation center starts its operation. Accordingly, two possible working shifts, namely 9 and 24 hours, are considered for the design of the scenarios. Like truck drivers, teleoperators cannot control a vehicle continuously for safety reasons. Looking at monitors for a long time is also not healthy for teleoperators. Since there are no available experiments to quantify this threshold, we defined the same maximum allowable teleoperation hours (4.5 hours) for the operators [37]. After this amount of teleoperation hours, the teleoperators must rest for 45 minutes. Teleoperators are also allowed to break down the rest into 10 minutes of short rest after each teleoperation.

## 4  Results and discussion

In this section, we present the results of our simulation study. As can be seen in Table 2, a combination of all the settings allowed us to experiment with 360 scenarios through which we could examine teleoperation under extreme and non-extreme system conditions. In this section, we summarize the most notable findings from these experiments. We begin with an extreme scenario in which the vehicle-to-teleoperator ratio is the lowest (0.3), the working shift of teleoperators is 9 hours, and the takeover time is 3 minutes. Next, we compare this scenario with the baseline (normal truck operation) for face validity. Then we explore the impact of various teleoperator-to-vehicle ratios and discuss a potentially optimal setting. Furthermore, we examine the impact of different working shifts and simulation start times on the optimal scenario. Finally, we apply a cost-benefit analysis to estimate the financial performance of the teleoperation for carriers.

### 4.1 Extreme teleoperation scenarios

In this scenario, the vehicle-to-teleoperator ratio is minimum, meaning that the number of teleoperators available in this system is only 30% of the number of vehicles. In this scenario, the takeover time is maximum since there is a lot of pressure on teleoperators, and hence higher time is needed for them to take over a truck. The working shift is 9 hours meaning that the teleoperation center can only give service to vehicles for 9 hours and the teleoperators start their work at 5:00 AM. This start time is aligned with the time that carriers often start their activities, and the duration will support the morning peak in teleoperator demand. We mark the end of the working time, which corresponds to the baseline makespan, by a vertical line to show the system state at this time but we continue the simulation until the completion of all tours in order to calculate delays and tour complication times under each scenario.

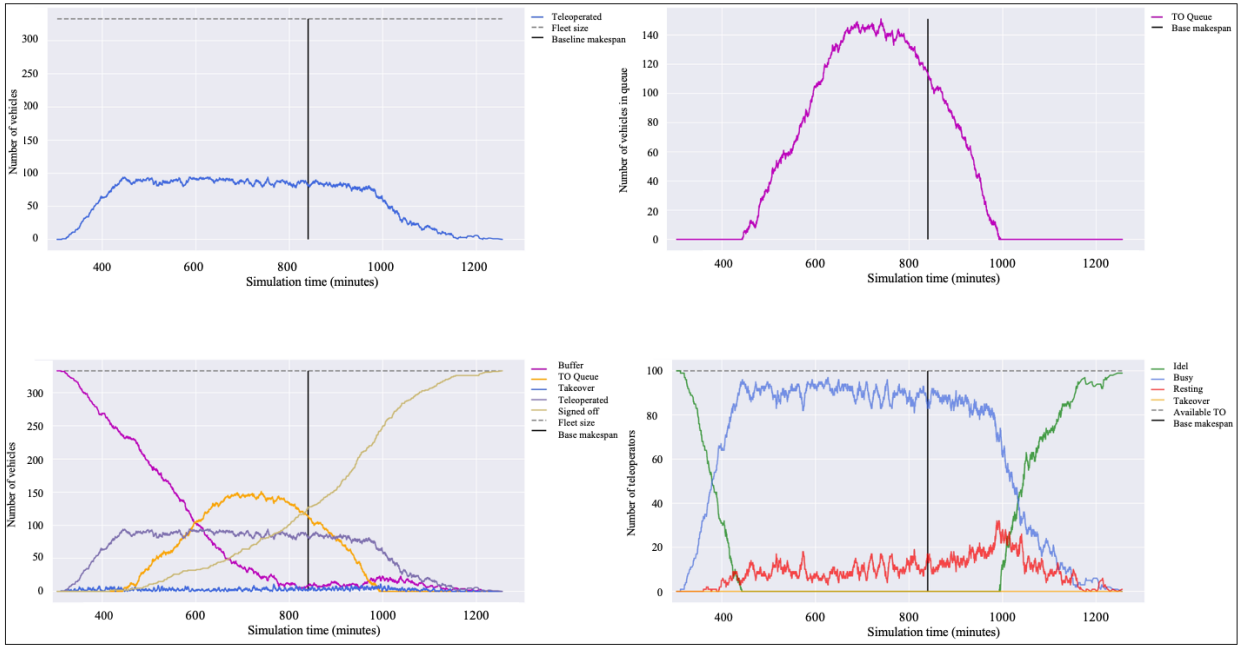

*Figure 6　Summary of teleoperation for the teleoperator-to-vehicle ratio of 0.3*

Figure 6 provides a comprehensive summary of the simulation scenario, depicting the number of vehicles engaged in different activities over the simulation timeline, along with the fleet size, the number of teleoperators who are either busy or idle at various points in time, the total number of available teleoperators, and the size of the teleoperator queue at each time interval. The figure offers a clear visualization of the key performance indicators and highlights the operational aspects of the system at various stages of the simulation.

The data shows that the queue size increases due to the limited number of available teleoperators. Additionally, the teleoperation scenario requires a longer duration to complete all tours compared to the baseline makespan. In other words, the distance between the vertical black line illustrated in Fig. 6 and the end of the simulation indicates the delay that teleoperation would pose to the system in a given scenario. This delay is attributed to the extended waiting time and takeover time, which teleoperation introduces to the system. Notably, a lower teleoperator-to-vehicle ratio would lead to longer waiting times, thus rendering the logistics system inefficient despite the reduction in teleoperator count. Hence, it is evident that the benefits of an exceptionally low teleoperator-to-vehicle ratio come at the cost of reduced efficiency and increased waiting times.

*Table 3  Summary of key performance indicators for the simulated scenario with teleoperator-to-vehicle ratio of 0.3*

|  | mean | std | min | 25% | 50% | 75% | max |
|---|---|---|---|---|---|---|---|
| AVG vehicle utilization ($Util_k$) | 0.144 | 0.0055 | 0.14 | 0.14 | 0.14 | 0.15 | 0.15 |
| AVG TO utilization ($Util_{to}$) | 0.724 | 0.0055 | 0.72 | 0.72 | 0.72 | 0.73 | 0.73 |
| AVG waiting time per vehicle ($wt_k$) | 142.292 | 8.8283 | 132.06 | 134.39 | 143.85 | 149.03 | 152.13 |
| AVG waiting time per vehicle in queue ($wt_Q$) | 84.724 | 3.8742 | 80.84 | 81.89 | 83.34 | 87.73 | 89.82 |
| MAX waiting time per vehicle in queue | 139.54 | 7.2045 | 133.68 | 136.67 | 137.28 | 137.98 | 152.09 |
| AVG queue length | 38.716 | 1.9177 | 36.27 | 37.37 | 39.26 | 39.52 | 41.16 |
| Max queue length | 146.8 | 5.9749 | 137 | 146 | 148 | 151 | 152 |
| AVG delay (D) | 0.8607 | 0.0366 | 0.8109 | 0.8421 | 0.8608 | 0.886 | 0.904 |

Table 3 shows the summary statistics of the key performance indicators for this simulated scenario over 5 replicas. Despite the desirably high teleoperator utilization rate of 72%, vehicle utilization is only 14%. Therefore, this scenario has the lowest level of service (indicated by the high waiting times in queues) as there are very few resources (teleoperators) to be assigned to each vehicle, and hence trucks must wait, on average, as long as 84.7 minutes in the queue for teleoperators. As we can see from Table 3, the freight transport system in this scenario will meet on average 86% delays in its operation, which imposes excessive costs on the system.

To represent the two ends of the spectrum in terms of the trade-offs between teleoperation labor cost and the level of service, we consider another extreme case where the teleoperator-to-vehicle ratio is equal to 1 and the takeover time is 0. The only difference between this scenario and common freight transportation is that each truck has one teleoperator instead of one truck driver. This scenario will not be beneficial for the business model of teleoperation since the objective of resolving the lack of truck drivers is not met. However, studying this scenario can provide face validity for the model and establish a baseline for evaluating the relative efficiency of other teleoperation scenarios.

In this scenario, the opposite phenomenon happens as compared to the previous extreme scenario (see Fig. 7). Since one operator is available for each truck, the queue length will always be zero. However, this comes at the cost of a low teleoperator utilization rate, thereby having many idle teleoperators at any point in time, which means higher labor costs. Moreover, as shown in Fig. 7, the number of busy teleoperators in this case never gets close to the number of available teleoperators (i.e., the instantaneous utilization rate of teleoperators never reaches 100%). This means a higher level of service (indicated by lower queue times) which comes at the price of higher labor costs due to a large number of teleoperators that are idling in the system.

By examining these two extreme scenarios on opposite ends of the experimental spectrum, we can conclude that there exists a trade-off between the cost of labor for teleoperators and the level of service provided. There is likely an ideal teleoperator-to-vehicle ratio that offers the optimal balance for this trade-off. In the following subsection, we will delve deeper into this topic to explore the potential trade-off.

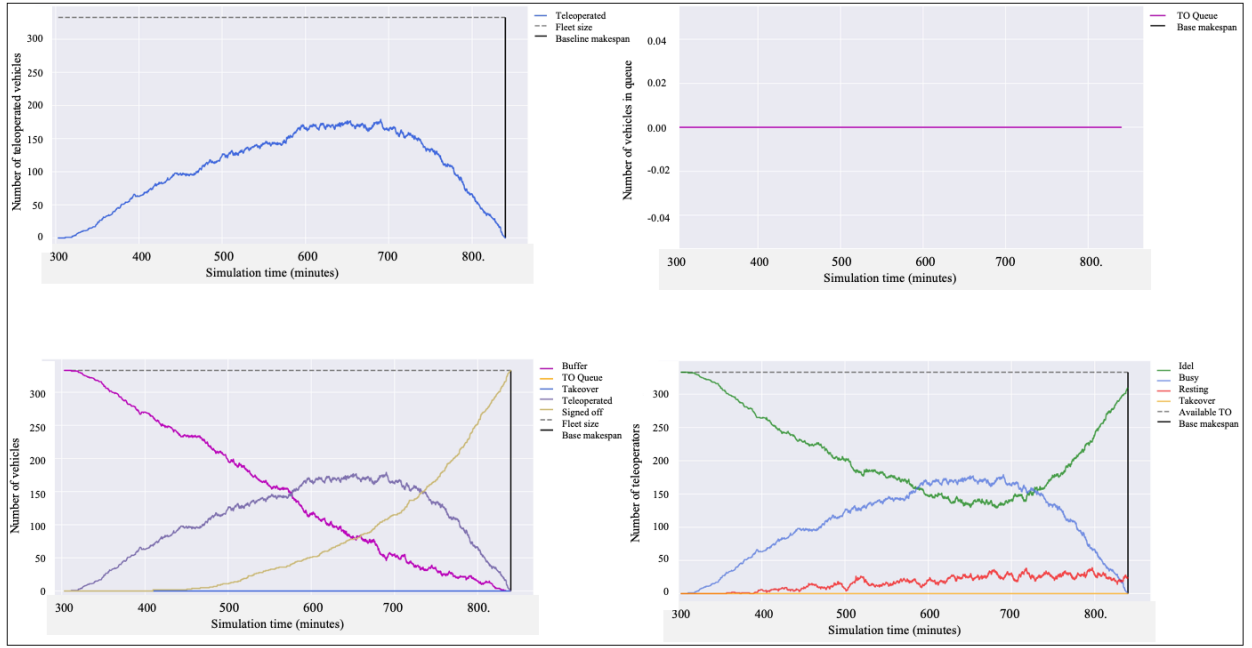

*Figure 7　Summary of simulation for the teleoperator-to-vehicle ratio of 1.0*

### 4.2　Optimal Teleoperator-to-vehicle ratio

In this section, we utilize different KPIs to discuss several factors that could impact the ideal ratio of teleoperators to vehicles. To this end, we explore the changes in the value of KPIs over different ranges of the teleoperator-to-vehicle ratio. These KPIs are tour completion, distance completion, delay rates, and average queue duration.

As evidenced by Figure 8, with a teleoperation-to-vehicle ratio equal to 0.6 and starting time 0:00, all tours are completed, the average distance completion rate is 1 and the queue duration and delay in the system is almost zero. This means with only 60% of current human resources we can drive trucks without any loss. It clearly shows the potential gain of teleoperation in increasing the efficiency of the system. We could quantify the gain from this efficiency as follows:

$$Gain\ Ratio = \frac{N_{base} - N_{To}}{N_{base}} \times 100 \qquad (9)$$

where, $N_{base}$ is the number of drivers in the base scenario (without teleoperation) and $N_{To}$ is the number of teleoperators in the teleoperation scenario. Using Equation 10, we could for the scenario with a teleoperation-to-vehicle ratio equal to 0.6, we could conclude that the teleoperation system is 40% more efficient in terms of labor cost.

We should note that the loss in the system increases exponentially when we reduce the teleoperator-to-vehicle ratio to less than 0.6. However, with a teleoperator-to-vehicle ratio equal to 0.5, more than 90% of tours are completed, 95% of tour distances are driven, and the delay in the system ranges between 5% to 10% depending on the magnitude of the takeover time (see figure 8). In this scenario, trucks have a waiting time of no more than 4 to 10 minutes on average. These waiting times and delays pose operational costs to the system which can, to some extent, be compensated by the labor cost savings. To quantify the gain of the system in such cases, we require a more complex cost structure associated with a certain level of service of the teleoperation which can be calculated as the ratio of the difference between the current cost of transport minus the cost of running the transport with the teleoperation.

$$Gain\ Ratio = \frac{N_{base} \times MS_{base} \times w_{base} - N_{To} \times MS_{To} \times w_{To}}{N_{base} \times MS_{base} \times w_{base}} \qquad (10)$$

where $MS_{base}$ is the simulation duration (makespan, see Equation 5) in the base scenario, $MS_{To}$ is the simulation duration or makespan in the teleoperation scenario, and $w_{base}$ and $w_{To}$ is the drivers' and teleoperators' hourly wage, respectively. Assuming equal hourly wages for the truck drivers and teleoperators, equation 8 can be simplified as follows:

$$Gain\ Ratio = \frac{N_{base} \times MS_{base} - N_{To} \times MS_{To}}{N_{base} \times MS_{base}} \times 100 \qquad (11)$$

Using the simulation model presented in this study and based on the results of the multiple simulation scenarios, the optimal teleoperator-to-vehicle ratio can be specified for any given level of service defined by any KPI (e.g., max queue duration or average delay). For instance, considering the average queue duration as a service level indicator, the lowest number for the teleoperator-to-vehicle ratio (the most economical option) can be specified for any desired average queue duration. The results depicted in Figure 8 show that to ensure an average wait time below, for example, 20 minutes, a teleoperator-to-vehicle ratio of 0.5 will suffice (see Fig. 8).

In this scenario, the makespan is 560.1859 and 598.1721 minutes for the baseline and teleoperation scenarios respectively. The number of truck drivers in the baseline is 250 and the number of teleoperators in the teleoperation scenario is 125 (To/Veh = 0.5). Using Equation 11, the gain of the system is 47% which exceeds the gain in the scenario with the 0.6 teleoperator-to-vehicle ratio (no delay scenario). This means that the lower labor costs associated with the lower teleoperator-to-vehicle can compensate for the higher labor costs associated with the higher makespan (delay).

Figure 8 also shows that a lower teleoperator-to-vehicle ratio can be achieved without any loss if the teleoperation starts at 8:00 as compared to the start time 0:00 or 5:00. By tolerating only 10 to 15 minutes of average queue time, one could complete more than 95% of the tour distances with a teleoperator-to-vehicle ratio of 0.45 within the baseline makespan. The average gain of this scenario is 49% . The reason we can achieve higher gain with a lower teleoperator-to-vehicle ratio in this scenario is that tours starting after 8 include more trips (legs) as compared to the tours starting earlier in the morning. Therefore, the buffer time between trips allows teleoperators to drive more vehicles. It should be noted that the higher takeover time can slightly reduce the gain. However, this loss is relatively very low and hence negligible. We also would like to bring to the attention that these gains are potentially achievable with only 1% teleoperation market penetration rate. For a higher teleoperation rate, we could expect much larger gains.

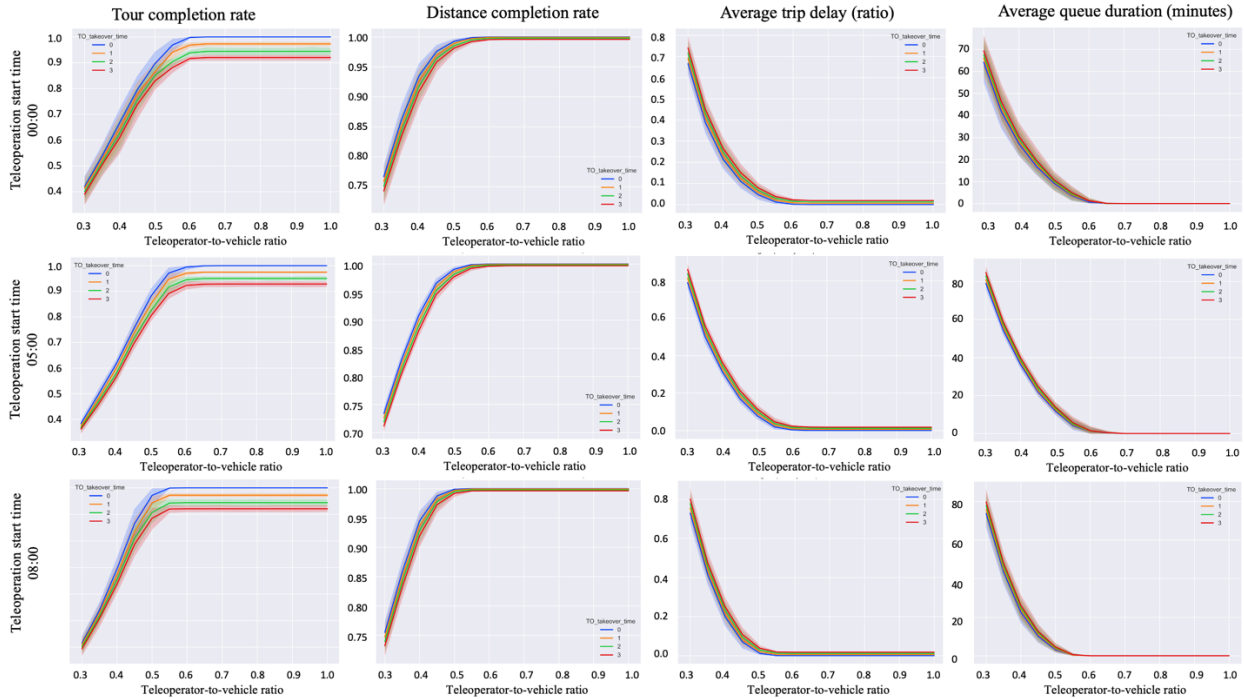

*Figure 8. KPIs versus teleoperator-to-vehicle ratio for different start time scenarios with 9 hours of working shift.*

We also explored the performance of the teleoperation for the 24-hour working shift. Figure 9 shows the KPIs for different start time scenarios with 24-hour working shifts of teleoperation. The most salient finding from this figure is that with the larger time span that the teleoperation system is active, a lower number of teleoperators is needed to drive all vehicles. Among all scenarios, the best teleoperation-to-vehicle ratio is 0.3 where the start time is 0:00 and the working hours of the teleoperation center is 24 hours. In this scenario, we have 250 teleoperators that drive 800 vehicles. It is evident that trucks must wait between 4 and 12 minutes on average to be assigned to a teleoperator which leads to a makespan of 1530.855 (the baseline makespan is 1472.246 minutes). The total gain of this system is 67% for the simulated day (using Equation 11). It should be noted that higher gains in 24-hour scenarios are partially due to the wider spread of peak times within the 24-hour working shift, more idle times within the system that affect all averages, and the higher number of teleoperators in the system due to having a higher number of tours in 24-hour work shifts.

We should bring to the attention that the gain calculation in this study is limited to labor costs only. One could argue that the longer makespan associated with a lower teleoperation-to-vehicle ratio may pose other costs like late delivery to the operators. Please note that a longer makespan does not necessarily mean late delivery. This mainly could mean that the same amount of delivery will be scheduled for a longer time window. We, therefore, may require different tour planning within the concept of teleoperation. However, we believe that unreliability in the teleoperated freight transport, that are attributed to queues and takeover times, can pose extra costs to the operators. In [38], authors calculated the value of reliability (VOR = 15 Euros/hour for 2-40 tone trucks) and the value of time (38 Euro/hour for 2-40 tone trucks) for freight transport in the Netherlands using collected stated-preferences data. In their research, the value of time (VOT) is associated with the expenses involved in offering transport services. If the duration of transportation decreases, it would free up vehicles and personnel for additional transports, leading to savings in vehicle and labor costs. The findings from the Netherlands and other countries suggest that the VOT linked to transport services aligns reasonably well with the hourly cost of vehicles and labor, particularly in the context of road transport [39]. Calculating the vehicle costs and reliability of transport in the teleoperation is beyond the scope of this research and hence ignored in favor of the teleoperation under a certain level of service thresholds that imposes delays. However, it is important to note that the absence of these calculations does not undermine the feasibility of teleoperation with a teleoperator-to-vehicle ratio of less than 1 and zero delays.

Moreover, in some cases, trucks waiting in queues may incur parking fees. Due to a lack of empirical data, these costs are challenging to measure in the study. Nevertheless, an interesting possibility could be the autonomous connection of trucks to nearby charging stations while waiting for a teleoperator in a queue. However, implementing such a scenario would necessitate a new planning and scheduling framework for teleoperation, which is beyond the scope of this study.

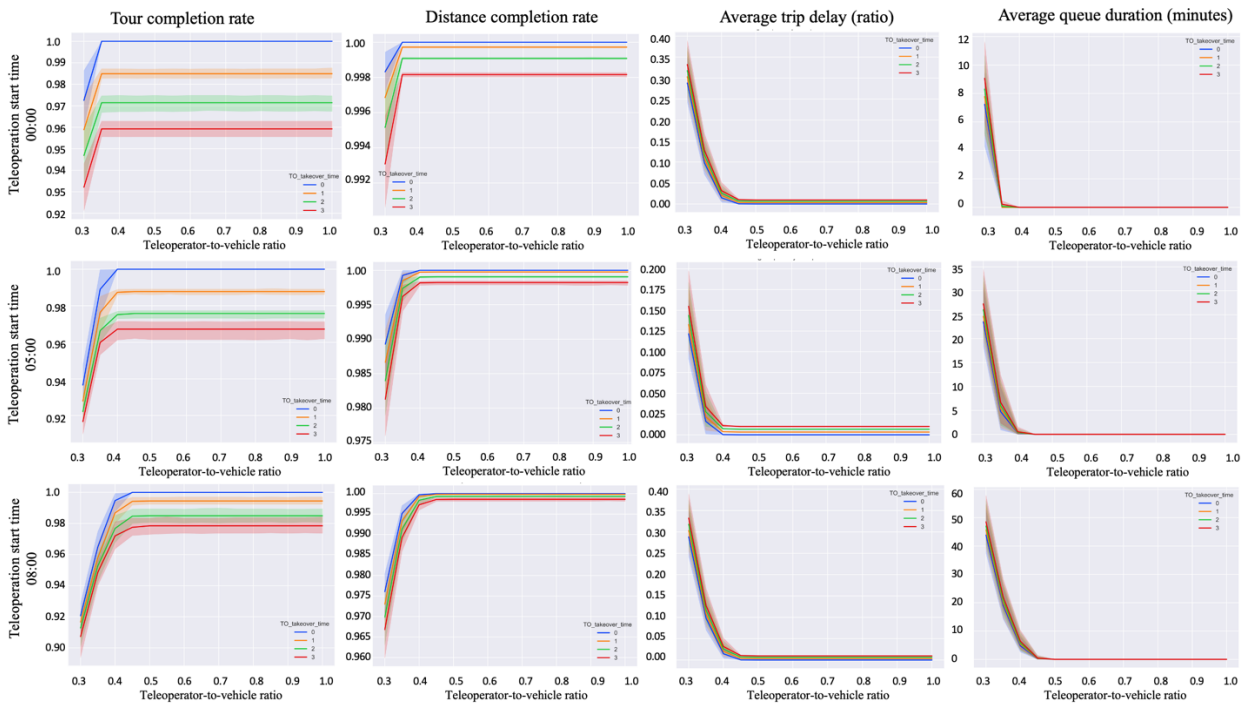

*Figure 9 KPIs versus teleoperator-to-vehicle ratio for different start time scenarios with 24 hours of working shift*

# 5 Conclusion and future research

The teleoperator-to-vehicle ratio is a crucial factor for the evaluation of the business case of teleoperated truck transport in logistics operations. The benefits of teleoperation increase when the capacity of teleoperators can be efficiently utilized by switching from an idle vehicle to another vehicle that is ready to move. This leads to higher utilization rates of teleoperators, thereby making it possible to operate a fleet of vehicles with teleoperator-to-vehicle ratios lower than one. However, when the number of teleoperators is lower than the number of vehicles, anytime a vehicle needs to move, it might need to wait in the queue before a teleoperator is assigned to it. This means there is a trade-off between the utilization rate of teleoperators, which represents the labor cost of teleoperated driving, and the waiting time for teleoperators, which represents the service level of teleoperated driving service.

In this study, we explored the trade-offs between the teleoperator-to-vehicle ratio and the level of service in logistics operations employing a simulation framework. Our simulation results indicated that the optimal teleoperator-to-vehicle ratio is dependent on the fleet size, operating hours, and the required level of service. We showcased the potential of our simulation framework for defining the optimal teleoperator-to-vehicle ratio given any service level defined by any KPI. By means of a case study that utilizes the freight transport trips within the region of South Holand in The Netherlands, we showed that with a teleoperation market penetration rate of 1%, a teleoperator-to-vehicle ratio of 0.5, and a standard 9-hour work shift, more than 95% of tour distances are completed and the system must only bear the burden of fewer than 10 minutes of waiting time on average. This confirms great promise for a positive business case for the concept of teleoperated driving as a service. In addition, we quantified the potential benefits of using teleoperation in our simulation scenarios.

We showed that tours (thereby demand for teleoperation) are not evenly distributed during working hours. In this study, we showed how to minimize teleoperator queue times by defining optimal teleoperator-to-vehicle ratios based on simulation results. An alternative approach for dealing with the peak demand issue is dynamic resource (teleoperator) allocation to maximize teleoperator utilization, which is a promising future research direction.

Two other interesting future research directions stem from considering the idea of 24-hour operation of teleoperation centers. First, the current tour planning/scheduling paradigm assumes that human drivers control trucks, and they have restrictions in terms of operating hours and conditions. However, with teleoperation centers providing continuous service around the clock for 24 hours, these restrictions can change or be eliminated. Fewer constraints for tour scheduling means better schedules. Future studies can leverage this benefit to propose new tour planning methods.

Another new possibility is coordinating teleoperation centers from different time zones. Since with teleoperated driving, there is no need for the vehicle and the teleoperator to be in the same time zone, teleoperation centers can be located in different time zones to provide 24/7 teleoperated driving service without the need for teleoperators to work overtime or outside the standard working hours. Such business models are currently in use for call centers, technical support, and customer service centers. This can lead to significant reductions in delivery times, but it requires a completely different planning paradigm. One that is not restrained by the driver's need for rest and sleep yet is still bound by the availability of the other actors in supply chains that interact with trucks and drivers (e.g., terminals, ports, storing and distribution facilities, end customers, etc.).

Another important topic related to teleoperated driving in logistics operations is the non-driving responsibilities of the drivers. Truck drivers are usually responsible for the vehicle, cargo, and communications with other supply chain actors as well as other road users and sometimes customers. Taking the driver out of the vehicle means these responsibilities need to be fulfilled by automation, digitalization, or other supply chain actors. A preliminary exploration of these responsibilities and viable solutions for overseeing them is provided in [40]. However, these are still open research questions and require further investigation.

## Acknowledgment

The Authors would like to thank Dr. Michiel de Bok from TU Delft for his help with MASS-GT simulation experiment for Zuid-Holland.